# Champagne Bubbles : Isolation and Characterization of amphiphilic macromolecules responsible for the stability of the collar at the Champagne / air interface

Zouleika Abdallah[1], , Véronique Aguié[2], Roger Douillard[2] and Christophe Bliard[1]

[1]FRE 2715 CNRS / URCA, Moulin de la  housse, Bât 18, BP 1039, 51097 Reims cedex 2, France.
[2]UMR 614 FARE INRA / URCA, 2 Esplanade R. Garros, BP 224 51686 Reims cedex 2 , France.

**Introduction** The effervescence and the stability of the ring of fine bubbles crowning the surface of a champagne glass, the "collar", constitute one of the hallmarks of Champagne. Defects in the stability of this collar are not well understood and account for a significant proportion of bottle return. This study aims to better understand the surface properties of champagne wine such that the foaming properties can be controlled more effectively. Early studies on Champagne foaming properties using the "Mosalux" measurement of the foam level formed by air flow in champagne through fritted glass pointed to a link between protein concentration and foam level [5] but no satisfactory correlation between protein content and foam stability was established. Later measurements were conducted with either ultra-filtrates or ultra-concentrates [4]. The authors demonstrated that macromolecule concentration was an essential parameter in the foam stability.

The stability of bubbles is usually ascribed to the presence of an adsorption layer formed at the gas/liquid interface and its properties [3]. Thus surface properties of Champagne were analysed by ellipsometry and tensiometry. Measurements conducted on base wine, on ultra-filtered base wines and degassed champagne samples showed the presence of an adsorption layer formed at the air/champagne wine interface [6] and that adsorption layer being composed of macromolecules in a $10^4$ to $10^5$ molecular range [7]. Previous studies on champagne wine macromolecules had shown wine macromolecules to be mostly proteins and polysaccharides [8] with very little insight as to the chemical constitution. The present study describes the isolation and characterization of these macromolecules and their link with the adsorption layer.

## Materials and methods

Champagne Samples
Wines samples were provided by the C.I.V.C. (Comité Interprofessionnel du Vin de Champagne, Epernay): from the three grape varieties: Pinot noir, Chardonnay and Meunier, from the 2004 harvest. A sample of the 1998 Pinot noir was also analysed.

Isolation protocol
Champagne base wine macromolecule isolation was performed in a stirred frontal ultra-filtration cell (fig 1) on a $10^4$ molecular weight cut-off (MWCO) regenerated cellulose membrane under 100KPa $CO_2$ pressure. Ultra-filtrate (UF, poor in macromolecules) and ultra-concentrate (UC rich in macromolecules) were collected and analysed.

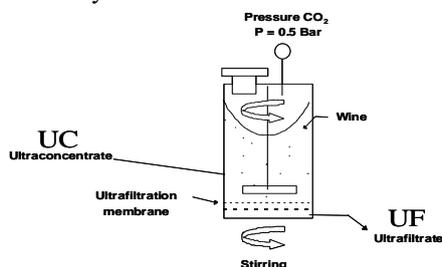

**Figure 1.** Schematic representation of the stirred frontal ultra-filtration cell

NMR Spectroscopy
$^1H$ experiments were recorded on a Bruker Advance 500 spectrometer, as $D_2O$ solution at 298 K in 5 mm o.d. tubes, calibrated on the water peak at 4.82 ppm.

Ellipsometry
Measurements were performed using a spectroscopic phase modulated ellipsometer (UVISEL, Jobin Yvon, Longjumeau, France). The ellipticity coefficient of the adsorption layer formed at the air/champagne interface was measured at the Brewster conditions, $\bar{\rho}_B$, for the substrate at 412 nm with an incidence angle of 53°5 according to the optical laws [1]. This study was carried out on reconstituted wine from ultra-concentrated fractions in the ultra-filtrate solution (UF) at the initial concentration in champagne.

Elementary analyses
Elementary analyses were performed on a Perkin Elmer CHN 2400 apparatus.

Analytical methods
Neutral sugars were analysed by GLC as their corresponding alditol acetates [9], using a SP 2380 capillary column (0,25mm) coupled to a H-P 3380 A integrator. Alditol acetates were obtained after three steps :
- $H_2SO_4$ hydrolysis to monosaccharides
- $NaBH_4$ reduction of monosaccharides
- Acetylation of alditols with $Ac_2O$

Ion-exchange chromatography[2]

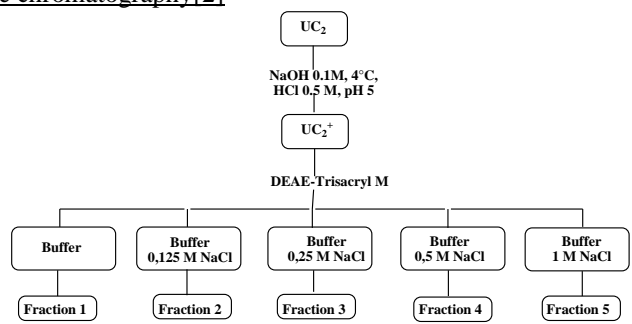

**Figure 2.** Example of fractionation procedure

Ultraconcentrates were saponified with 0.1M NaOH (overnight, 4°C). The solution was then acidified to pH 5 by addition of 0.5M HCl solution and extensively dialysed against distilled water and freeze-dried in $H^+$ form. Thereafter the sample (50 mg) was suspended in 10 mL of 0.05M phosphate buffer (pH=6.3) and the solution was loaded onto a (10x200mm) DEAE-Trisacryl M column, previously equilibrated with the same buffer, then eluted at 30 mL/h flow rate. The column was eluted with 30 mL buffer and then successively with the buffer containing respectively, 0.125, 0.25, 0.5, 1M NaCl, 30 mL each and finally by 0.5M NaOH. The fractions were dialysed against distilled water and freeze-dried.

Size-exclusion chromatography
Molecular weight distributions were studied by HPSEC by applying 100 µl samples of 2mg/mL of the previously obtained fractions onto two serial Shodex-OHpak B-805 (7.5x300mm) then Shodex-OH-pak B-804 (7.5x300mm) columns, equilibrated at 0.3 mL/min in a 0.05M $NaNO_3$ solution. The column effluent was monitored using a IOTA 2 (Precision Instruments, Marseille) refractive index detector. The weight average molecular weight was calibrated with 10 KDa to 2MDa dextrans standards.

## Results and discussion

Setting up the macromolecule isolation protocol
In order to set up the macromolecule isolation protocol, two ultra-filtrations, carried out on Pinot noir 1998 champagne, led to:
- **$UC_1$** (raw product) corresponding to 7.5 mL ultra-concentrate, freeze-dried. 165,2 mg were recovered as a brown solid, giving an initial concentration of 0.220 g/L.
- **$UC_2$** (purified product) corresponding to 7.5 mL ultra-concentrate, dia-filtrated several times (5x50mL distilled water) and freeze-dried. 104 mg were recovered as a greyish solid, giving an initial concentration in the bottle of 0.139 g/L.
These ultra-concentrates were submitted to NMR studies in order to investigate their chemical nature.

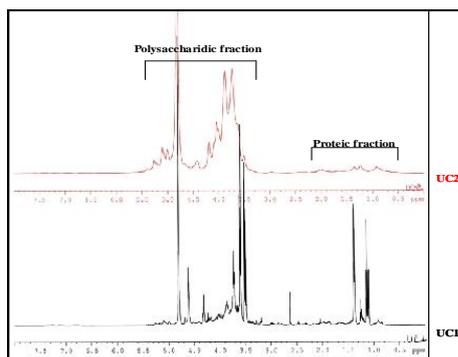

**Figure 3.** NMR spectra of **$UC_1$** and of **$UC_2$**

Proton NMR analysis showed
- for **$UC_1$** to be mainly composed of a polysaccharidic fraction and organic acids such as lactic and tartaric acid

- for **UC₂** to be mainly composed of a polysaccharidic fraction (spectral characteristic zone: 6-3 ppm) with a minor protein fraction (spectral zone: 2,1-0,5 ppm). The other products with lower molecular mass, such as organic acids were not present after the washing step.

Then these ultra-concentrates were submitted to surface properties studies using ellipsometry (Fig 5). It was found that the ellipticity of UF sample is positive and close to +0.001, indicating the lack of an adsorption layer at interface. On the contrary, in the case of the reconstituted samples (UC1 and UC2), kinetic decrease and negative value of the ellipticity were indicative of the progressive formation of an adsorption layer at the air/liquid interface. Both fractions presented the same behaviour as champagne and the difference between them is not significant. Moreover, the fluctuations of the signal show that the layer is inhomogeneous. Hence they were considered as containing the components responsible for the formation of adsorption layer in champagne. Therefore the second method including the purification step was preferred in subsequent experiments.

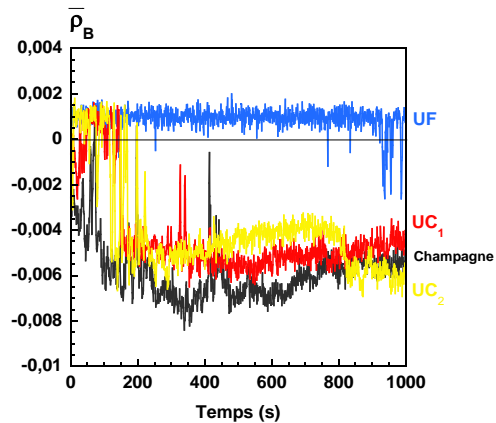

**Figure 5.** Brewster ellipticity, $\overline{\rho}_B$ vs time

Composition and physico-chemical characteristics of the three 2004 wine samples
Macromolecules content and elementary analysis: the results are summarized in the following table :

|  | Macromolecules concentration (mg/L) | Elementary analysis | | | proteic %= N% x 6.25 |
|---|---|---|---|---|---|
|  |  | % C | % H | % N |  |
| Pinot noir | 387 | 41.89% | 6.89% | 4.97% | 31% |
| Chardonnay | 375 | 43.03% | 6.63% | 4.91% | 31% |
| Meunier | 509 | 43.61% | 6.73% | 5.53% | 35% |

The macromolecules content was between 300 and 500 mg/L. The elementary analysis, in comparison with galactomannan from locust bean gum (%C: 43,63%; %H: 6.74 %;), confirmed the presence of glycosyl /polysaccharidic structure. The protein content was calculated as a function of nitrogen percentage, as determined by elementary analysis. It was found to be in the order of thirty % for Pinot noir and Chardonnay to thirty five % for Meunier. Proton NMR showed the three ultra-concentrates to have similar composition. Neutral sugar GLC analysis showed that these ultra-concentrates contained mainly mannose, galactose and arabinose.

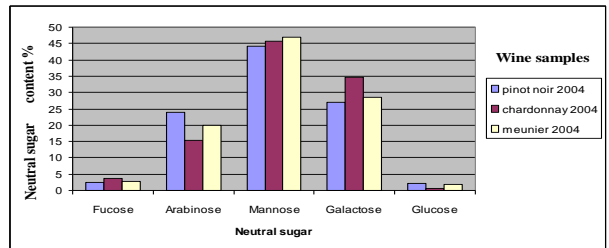

**figure 6.** Ultraconcentrates composition

Carbon NMR spectra showed a signal in 170 ppm characteristic of the presence of carboxylic acid. In the former sugar composition analysis the presence of uronic acids could not be analysed, the acid hydrolysis between of uronic acid unit requiring harsher conditions or would require prior reduction of the carboxyl group.

$UC_2$ Fractionation
In order to set the fractionation protocol a 1998 Pinot noir champagne UF, was used. The proportion of recovered sample is represented in the following diagram.

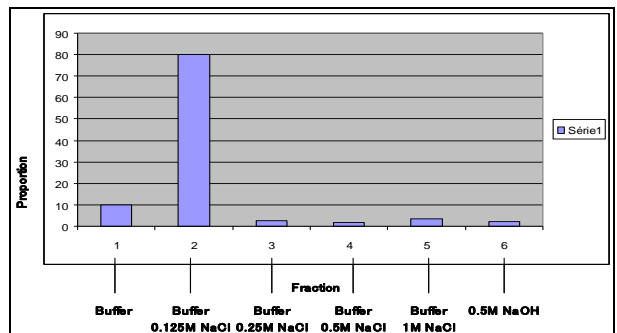

**Figure 7.** Quantitative results of fractionation

Ninety per cent of the sample were colleted in the first two fractions and 10 per cent in the following fraction. NMR showed Fraction 2 and 3 to be identical as well as for the three fractions 4, 5 and 6.
Fraction 1: NMR analysis, showed fraction 1 to be of proteic nature. Size-exclusion chromatography showed a narrow peak at a MW slightly lower than 10 KDa.
Fraction 2: NMR analysis showed it to be mainly composed of a polysaccharidic fraction with a minor protein fraction. SEC analysis showed four peaks ranging between 90KDa and 9KDa.
Fraction 6: NMR analysis showed fraction 1 to be of proteic nature with the presence aromatic component. SEC showed four peaks between 5KDa and 6KDa.

## Conclusion
The primary results obtained are
- the macromolecule isolation protocol by ultra-filtration of wine at a $10^4$ MWCO.
- the demonstration of the adsorption layer formation at the air / reconstituted wine (macromolecular fraction solution) by ellipsometry.
- The confirmation of a polysaccharidic nature with a minor protein fraction of this macromolecular fraction by NMR analysis.
-The Mannose Arabinose Galactose constitution chemical analysis.

- The isolation of three fractions by ion-exchange chromatography: one polysaccharidic fraction, two protein fractions, one of them containing aromatic residues.

*Acknowledgements*: We thank Dr R. Marchal from the laboratoire d'oenologie de l'Université de Reims Champagne Ardennes for technical help. This work has been supported by a Europol'Agro grant within the VINEAL project.